\documentclass[aps,twocolumn,amssymb,floats,floatflt,prb,amsmath,superscriptaddress,showpacs]{revtex4-1}
\usepackage{graphicx}

\begin{document}

\title
  {On the microscopic origin of the magneto-electronic phase separation \\*	in Sr doped {LaCoO$_3$}} 

\author{Zolt\'an N\'emeth}
\affiliation{Wigner Research Centre for Physics, Hungarian Academy of Sciences, H-1525 Budapest, P.O.B. 49., Hungary}
\affiliation{Department of Nuclear Chemistry, E\"otv\"os Lor\'and University, 
	 	P\'azm\'any s\'et\'any 1/A, H-1118 Budapest, Hungary}
\author{Andr\'as Szab\'o}
\affiliation{Wigner Research Centre for Physics, Hungarian Academy of Sciences, H-1525 Budapest, P.O.B. 49., Hungary}
\affiliation{Institute of Physics, University of Szeged, D\'om t\'er 9., H-6720 Szeged, Hungary}
\author{Karel~Kn\'{\i}\v{z}ek}
\affiliation{Institute of Physics ASCR, Cukrovarnick\'a 10, 162 00 Prague 6, Czech Republic}
\author{Marcin Sikora}
\affiliation{AGH University of Science and Technology, Faculty of Physics and Applied Computer Science, Solid State Physics Department, Av. Mickiewicza 30, 30-059~Krakow, Poland}
\author{Roman Chernikov}
\affiliation{HASYLAB, DESY, D-22607 Hamburg, Germany}
\author{Norbert Sas}
\affiliation{Wigner Research Centre for Physics, Hungarian Academy of Sciences, H-1525 Budapest, P.O.B. 49., Hungary}
\author{Csilla Bogd\'an}
\affiliation{Wigner Research Centre for Physics, Hungarian Academy of Sciences, H-1525 Budapest, P.O.B. 49., Hungary}
\author{D\'enes Lajos Nagy}
\affiliation{Wigner Research Centre for Physics, Hungarian Academy of Sciences, H-1525 Budapest, P.O.B. 49., Hungary}
\author{Gy\"orgy Vank\'o}
\affiliation{Wigner Research Centre for Physics, Hungarian Academy of Sciences, H-1525 Budapest, P.O.B. 49., Hungary}

\begin{abstract}
The nanoscopic magneto-electronic phase separation in doped La$_{1-x}$Sr$_x$CoO$_3$ perovskites was studied with local probes. The phase separation is directly observed by M\"ossbauer spectroscopy in the studied doping range of 0.05 $\leq$ $x$ $\leq$ 0.25 both at room-temperature as well as in the low-temperature magnetic phase. Extended with current synchrotron based X-ray spectroscopies, these data help to characterize the volume as well as the local electric and magnetic properties of the distinct phases. A simple model based on a random distribution of the doping Sr ions describes well both the evolution of the separated phases as well as the variation of the Co spin state. The experiments suggest that Sr doping initiates small droplets and a high degree of doping driven cobalt spin-state transition, while the Sr-free second phase vanishes rapidly with increasing Sr content.\end{abstract}

\date{\today}


\maketitle

\section{Introduction}

\begin{figure}  
\includegraphics[width=8.5cm]{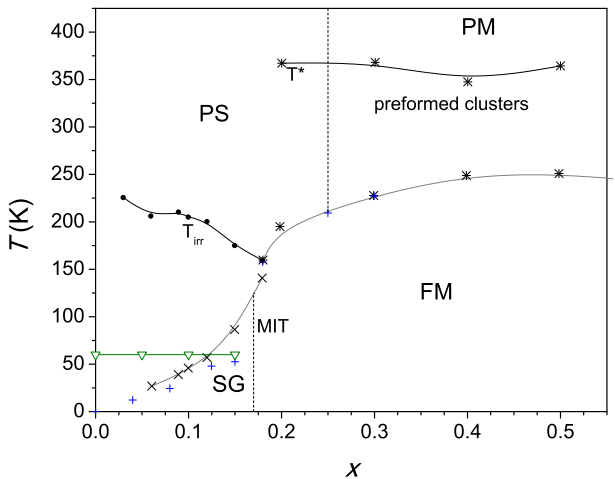}
\caption{\label{fig1} (Color online). Phase diagram of La$_{1-x}$Sr$_x$CoO$_3$ taken from Refs. \onlinecite{he2007,wu2003} (black dots, crosses, and asterisks), Ref. \onlinecite{kriener2004} (blue plus signs), and Ref. \onlinecite{smith2012} (green triangles). PS: paramagnetic semiconductor, SG: spin-glass semiconductor, PM: paramagnetic metal, FM: ferromagnetic metal, MIT: metal to insulator/semiconductor transition. Empty triangles represent spin state transition temperature from local NMR results \cite{smith2012}, $T_{irr}$ is the magnetic irreversibility temperature,\cite{wu2003} $T^*$ is the temperature where Sr rich clusters are expected to dissolve.\cite{he2009epl}}
\end{figure}

Doping-fluctuation driven magneto-electronic phase separation (MEPS), the spatial coexistence of multiple electronic and magnetic phases even in the absence of chemical segregation, is ubiquitous in complex oxides such as cuprates, manganites and cobaltates, where the phases separate into nanoscopic clusters with stable ferromagnetic (FM) order in a non-FM matrix.\cite{he2009epl} These nanoscale phases, which differ starkly in both conductivity and magnetic order, compete in an unexpectedly complex manner as a function of temperature, external and chemical pressure as well as hole doping. Moreover, the resultant cooperativity between the electronic and magnetic states is a remarkable feature that can help to advance developments in spintronics. MEPS is expected to occur in the perovskite type oxides La$_{1-x}$Ca$_x$MnO$_3$ or La$_{1-x}$Sr$_x$CoO$_3$ with compositions $x<0.5$, that exhibit not only colossal magnetoresistive (CMR) effects (developed by spin-dependent transport between isolated clusters), but also an exceptionally rich phase diagram as a function of temperature, external and chemical pressure, hole doping $etc.$ (Fig. \ref{fig1}).\cite{dagotto2001,he2009epl,wu2003,kriener2004,aarbogh2006}

While the crystal structure of LaCoO$_3$ has a rhombohedral symmetry (space group  R$\bar{3}$c), replacing some of the lanthanum ions with strontium reduces the rhombohedral distortion. This effect increases with doping $x$ level, for $x$=0.5 the structure is cubic at room temperature. In addition to the structural changes, Sr-doping causes variations in the charge and spin state of the Co ions leading to large variations in the magnetic and transport properties. In the case of the parent compound LaCoO$_3$, the low-spin (LS) state (t$_{2g}^6$e$_{g}^0$, S=0) of cobalt ions is favored at $T$=0, due to the slight dominance of $E_{CF}$ (crystal field splitting energy) over $E_{J}$ (Hund’s exchange energy) resulting in a nonmagnetic insulating ground state;\cite{ivanova2009,zhang2012prb} however, thermal excitation leads to a finite spin state. It is controversial whether the resulting paramagnetism is due to an intermediate-spin (IS) (t$_{2g}^5$e$_{g}^1$, S=1) or a high-spin (HS) (t$_{2g}^4$e$_{g}^2$, S=2) state of the Co$^{3+}$, or a mixture of the two.\cite{vanko2006prb,haverkort2006,klie2007} Doping also affects the magnetic properties rather strongly even at very low substitution levels: it has been reported that a single Sr$^{2+}$ ion induces a total spin change of $\Delta$$S$=10-16.\cite{yamaguchi1996prbSCO} The origin of these spin polarons is ambiguous, although their existence has been confirmed in separate experiments.\cite{podlesnyak2011,PhysRevLett.94.037201,phelan2006,smith2008prb} In addition to the spin state variations, doping the system with divalent Sr$^{2+}$ (and thus replacing the trivalent La$^{3+}$) is supposed to introduce electron holes in the CoO sublattice either by the oxidation of Co$^{3+}$ to Co$^{4+}$ or creating electron holes on the oxygen. The interplay of the electronic, spin and lattice degrees of freedom leads to a rich phase diagram in these perovskites, which involves semiconductor (SC) to metal as well as spin glass (SG) to unconventional ferromagnet (FM) transitions, which can be interpreted in terms of the MEPS model.\cite{he2009epl,podlesnyak2011} It is generally accepted that the ferromagnetism in the hole-doped La$_{1-x}$Sr$_x$CoO$_3$ arises as a result of the double exchange (DE) interaction between nominally either Co$^{3+}$(IS) and Co$^{4+}$(LS) or Co$^{3+}$(HS) and Co$^{4+}$(IS) ions. Moreover, the FM state evolves by increasing the doping level $x$.\cite{wu2003} However, as there is still a serious ambiguity concerning the electron and the spin-density at the cobalt ions, the microscopic origin of the large variations in transport properties and magnetic ordering, as well as the nanoscale cluster formation is far from being understood.

We decided to revisit the La$_{1-x}$Sr$_x$CoO$_3$ (LSCO) system with the help of local probes from both synchrotron and laboratory techniques, which can verify the structure, determine the local charge and spin states as well as the extent of the magnetic ordering in the temperature and doping window of interest. Techniques including synchrotron X-ray spectroscopies and M\"ossbauer spectroscopy can provide data on local valence electron density and distribution as well as local magnetic ordering, thus address the pending questions on spin transitions and cluster formation.

M\"ossbauer spectroscopy is a technique with several unique characteristics that make it an appropriate tool to investigate coexistent nanoscale magnetic phases. It is an element-specific probe, which is very sensitive to magnetism. It is a local probe that shows the interactions of a single nucleus with the surrounding charges and magnetic moments; therefore, charge or spin-density variations on the nanometer or larger scales do not cause averaging or smearing the signal. Different atomic environments are reflected by the different values of hyperfine interactions including electric monopole, electric quadrupole, and magnetic dipole interactions. It is possible to disentangle parameters that are characteristic of different sites, making this technique well-suited for phase analysis. Furthermore, the mother nuclide for the $^{57}$Fe M\"ossbauer spectroscopy is the $^{57}$Co, which can easily be introduced into the cobalt sites of the LSCO system. Complications may arise in insulators as the nuclear decay can cause chemical after-effects;\cite{spiering1990} however, the conductivity of the cobaltates studied is sufficient to settle any after-effects well before the M\"ossbauer-window opens (\textit{i.e.}, a few nanoseconds after the electron capture of $^{57}$Co).

Despite the above-described capabilities of M\"ossbauer spectroscopy, until recently only pioneering work has been conducted on LSCO perovskites in the 1970's. V. G. Bhide \textit{et al.} proved the feasibility of emission M\"ossbauer spectroscopy on these cobaltates and explained the results on the basis of itinerant-electron ferromagnetism \cite{bhide1972prb,bhide1975prb} outlined in the series of Goodenough's work.\cite{goodenough1966,goodenough1968jap} In the light of the recent results on the importance of the nanoscale magnetism, reopening the M\"ossbauer spectroscopy investigations of this system is expected to deliver new insights.

Synchrotron-based X-ray spectroscopies, including X-ray absorption (both X-ray absorption near edge structure (XANES), extended X-ray absorption fine structure (EXAFS)) and emission (XES) techniques, provide element selective information on local charge, spin and structural properties (see \textit{e.g.} Ref. \onlinecite{glatzel2005ccr}).

These data can help to discover the complex electronic and magnetic features of doped LaCoO$_3$ and already provided a handful of important data on these perovskites.\cite{vanko2006prb,haverkort2006,vanko2006jpcb,sterbinsky2012,jiang2009,moodenbaugh2000prb,toulemonde2001,haas2004,lengsdorf2007}
To investigate the microscopic origin of the MEPS phenomenon, we explored systematically the local electronic and magnetic properties of the separated phases in the La$_{1-x}$Sr$_x$CoO$_3$ ($x<0.5$) perovskite series around its phase transitions in greater detail with the above local probes.

\section{Experimental section}

The samples La$_{1-x}$Sr$_x$CoO$_3$ ($x$ = 0, 0.05, 0.1, 0.15, 0.25, 0.5) were prepared by a solid-state reaction from stoichiometric amounts of La$_2$O$_3$, SrCO$_3$ and Co$_2$O$_3$. The precursor powders were mixed, pressed in the form of pellets and sintered at 1100$^\circ$C for 100 hours in oxygen flow. Characterization with X-ray powder diffraction (XRD) was performed with a Bruker D8 diffractometer using Cu~K$\alpha$ radiation, equipped with SOL-X energy dispersive detector. Their analysis proved the formation of a single phase for all $x$. XRD has also indicated that the additional heat treatments did not change the structure or phase composition of the material.

X-ray absorption data at the Co~K edge were collected on the beamline C of the Hasylab DORIS III synchrotron in transmission geometry. While monitoring the intensities $I_0$ and $I$ before and after the sample, for the sake of correcting possible instrumental energy shifts the absorption of a thin Co foil was recorded simultaneously, as well. The EXAFS data were reduced and background-corrected using the Athena software package.\cite{Ravel2005jsr} The experimental absorption edge position ($E_0$) was defined as the energy of the inflection point on the Co~K edge. EXAFS data of bulk La$_{1-x}$Sr$_x$CoO$_3$ samples with $x$ = 0, 0.05, 0.15, 0.25, 0.5 was investigated at 20~K and 300~K. In all fits, Fourier transformation and fitting ranges of $k$ and $R$ were 4-13~\AA$^{-1}$, and 1-3~\AA, respectively, using Hann windows. The first peak around 2~\AA, which corresponds to the first coordination shell, was fit to a theoretical Co-O EXAFS path generated by FEFF 6.\cite{rehr2000rmp}

X-ray emission spectra were recorded at the ID16 beamline of the European Synchrotron Radiation Facility (ESRF). A setup with a spherically bent Si(533) analyzer crystal placed on a 1-m diameter Rowland circle was used to record main K$\beta_{1,3}$ line and its K$\beta$' low-energy satellite lines of Co. All spectra were normalized to the beam monitor as well as to the spectral area. The spectra were aligned to have the same first momentum. In order to extract the average spin state ($S_{avg}$) of cobalt ions, the IAD (integral of the absolute values of the difference spectra) method was used.\cite{vanko2006jpcb} For low-spin Co$^{3+}$ reference, the spectrum of EuCoO$_3$ at 10~K was taken, which was subtracted from all spectra and the area of the absolute values of these differences was calculated. The resultant IAD values were converted to $S_{avg}$ according to the calibration published in Ref. \onlinecite{vanko2006prb}.

$^{57}$Co doping of the perovskites for the M\"ossbauer measurements was performed as follows. Aqueous solution of $^{57}$CoCl$_2$ in 0.1~M HCl was let evaporate to dryness at gentle heating (under infrared lamp). Then, stochiometric quantity (1:2 metal-to-ligand ratio) of ammonium citrate was added as aqueous solution of ammonia-ammonium citrate buffer (pH$\simeq$7). 60~$\mu$l of this solution with an activity of 80 to 110~MBq was added to each perovskite pellet, and it was dried under an infrared lamp. The samples were then subjected to a diffusion heat treatment for 2 hours at 1000$^\circ$C in O$_2$ atmosphere followed by cooling to room temperature in the furnace in the same constant O$_2$-flow.

For the emission M\"ossbauer measurements conventional constant acceleration type spectrometers (Wissel and~KFKI) were used. A standard $^{57}$Fe-enriched potassium hexacyanoferrate (PFC) absorber moving relative to the sample absorbed the $\gamma$-rays. Low temperature measurements were carried out using a through-flow liquid N$_2$ cryostat down to 77~K. Isomer shifts throughout this paper are given relative to $\alpha$-Fe at 295~K with sign convention used in transmission measurements. The velocity scales were calibrated using a $^{57}$Co-doped $\alpha$-Fe foil as a source. The M\"ossbauer spectra were analyzed with the MossWinn 4.0 program.\cite{klencsar1997}

\section{Results}

The X-ray powder diffraction patterns confirm that the samples contain only a single perovskite phase which is in good agreement with previous studies.\cite{Mineshige1996jssc} The structural parameters within the  R$\bar{3}$c space group were determined with the help of Rietveld analysis using the FULLPROF program; the resulting rhombohedral lattice parameters are listed in Table \ref{table1}. The rhombohedral distortion is reduced by the Sr doping, and no change in lattice symmetry was observed at room temperature in the doping range studied. The lattice parameters of the samples are in good agreement with those reported by Mineshige \textit{et al.} \cite{Mineshige1996jssc} The rhombohedral angle ($\alpha_r$) decreased almost linearly with $x$ showing that strontium doping decreases the rhombohedral distortion in the perovskite structure, thus increasing the symmetry of the unit cell.

\begin{table*}
\caption{\label{table1} Lattice parameters and calculated structural data of La$_{1-x}$Sr$_x$CoO$_3$ perovskites obtained with XRD and EXAFS.}
\begin{ruledtabular}
\begin{tabular}{lllllll}
& \multicolumn{4}{c}{XRD}& EXAFS & EXAFS \\
& \multicolumn{4}{c}{(300~K)}	& (300~K)	& (20~K) \\ \cline{2-5}	
$x$ & $a_r$ (\AA) & $\alpha_r$ ($^\circ$) & Co-O (\AA) & Co-O-Co ($^\circ$) & Co-O (\AA) & Co-O (\AA)
\\ \hline

0.05 & 5.386(1) & 60.73(1) & 1.932(1) & 164.5(3) & 1.923(9) & 1.92(1) \\
0.15 & 5.398(1) & 60.60(1) & 1.931(2) & 165.7(4) & 1.925(9) & 1.92(1) \\
0.25 & 5.405(1) & 60.45(1) & 1.929(1) & 167.4(3) & 1.922(9) & 1.93(1) \\
0.5 &  &  &  &  & 1.92(1) & 1.92(2) \\
\end{tabular}
\end{ruledtabular}
\end{table*}

\begin{figure}  
\includegraphics[width=8.5cm]{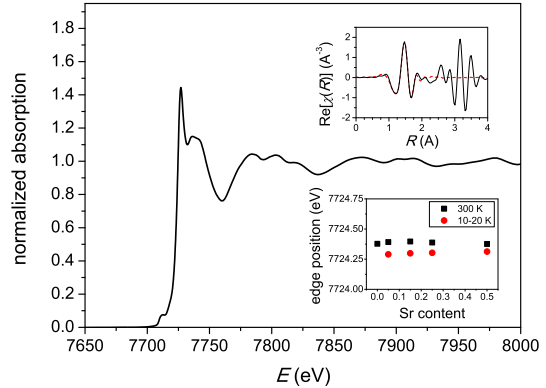}
\caption{\label{fig2} (Color online). Normalized X-ray absorption spectrum of the La$_{0.85}$Sr$_{0.15}$CoO$_3$ perovskite at 20~K. The upper inset shows the real part of the derived $R$-space EXAFS data (black solid line) fitted for the first coordination sphere (dashed red line). The lower inset plots the Co~K edge position \textit{vs.} Sr doping at two temperatures deduced from the first derivative of the XANES spectra.}
\end{figure}

The local structure around an ion can be determined element selectively by measuring extended X-ray absorption fine structure (EXAFS). EXAFS was recorded at the Co~K edge on a series of samples with Sr dopings $x$ = 0.05, 0.15, 0.25 and 0.5 at temperatures $T$ = 300~K and 20~K (Fig. \ref{fig2}). In accordance with the work of Jiang \textit{et al.},\cite{jiang2009} the spectra show one clear feature below 3~\AA\ in the R-space plot, corresponding to the Co-O bond of the CoO$_6$ octahedra. The fits show that both at room temperature and at 20~K the value of the Co-O bond length and also the width of the pair distribution function (showing no sign for a possible Jahn-Teller type distortion of the Co-O octahedron) do not change with Sr doping, thus the local structure does not show significant variations.
 
X-ray absorption is also used to study the electronic structure as it probes the unoccupied electron states. The X-ray absorption near-edge structure (XANES) is often considered as a probe of the valence state, so that the edge shifts are interpreted as signs of valence-state variations. The XANES of the studied LSCO systems agrees with those reported earlier.\cite{jiang2009} To make an attempt to determine the variations of the valence state of the cobalt ions, the inflection point of the Co~K edge data was determined. In agreement with previous studies, no significant change in the edge position was found either with Sr doping or with temperature, being $E_0$(300~K) = 7724.39$\pm$0.01 eV and $E_0$(20~K) = 7724.30$\pm$0.01 eV for all $x$. The lack of variations in the edge position, however, does not exclude possible electron-density differences, as the edge position is more dependent on the local geometry; therefore, the valence state cannot always be inferred from the XANES spectra.\cite{glatzel2009xafs14}

\begin{figure}  
\includegraphics[width=8.5cm]{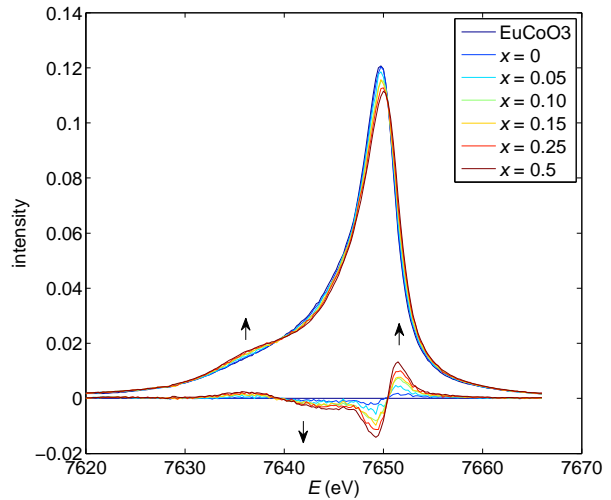}
\caption{\label{fig3} (Color online). X-ray emission spectra of the La$_{1-x}$Sr$_x$CoO$_3$ samples at temperatures 10-20~K (depending on the sample). The differences relative to the low-spin standard (EuCoO$_3$ at 10~K) are also plotted.}
\end{figure}

As the magnetic interactions in the LSCO perovskites depend primarily on the spin of cobalt ions, mapping the spin-state variations of cobalt is crucial. While bulk magnetization measurement techniques have been extensively used to investigate this issue, the local X-ray emission spectroscopy method can give element specific insights into the temperature and doping dependency of the spin state of Co. In agreement with the results obtained previously on LaCoO$_3$,\cite{vanko2006prb} the recorded XES spectra of the La$_{1-x}$Sr$_x$CoO$_3$ ($x$ =0, 0.05, 0.1, 0.15, 0.25, 0.5) samples show a gradual increase of the low-energy satellite of the main K$\beta$ line with both temperature and strontium doping level (an example at low temperature is shown in Fig. \ref{fig3}).
 
In our recent preliminary transmission and emission M\"ossbauer experiments on strontium-doped lanthanum cobaltates it was shown that iron at low concentrations (either directly doped or formed from $^{57}$Co by nuclear decay) is accommodated in the cobalt site of the lattice as high-spin Fe$^{3+}$ ion.\cite{nemeth2005epjb,nemeth2007} Similarly to the low-spin Co$^{3+}$, this ion in an octahedral environment is not Jahn-Teller active, so it has identical bond lengths, and also, its 3d electrons are symmetrically distributed, \textit{i.e.}, they do not contribute to the electric-field gradient (EFG) at the nucleus. Therefore, it can probe its electronic and magnetic environment.\cite{nemeth2007} The $^{57}$Co emission M\"ossbauer spectra of La$_{0.8}$Sr$_{0.2}$CoO$_3$ showed the coexistence of magnetic and paramagnetic subspectra below the magnetic transition temperature, $T_C$, in very good agreement with the MEPS model. Based on the isomer shift parameter, a greater 3d electron delocalization was observed in the magnetic phase, which could be tentatively associated with higher conductivity.

\begin{figure}  
\includegraphics[width=8.5cm]{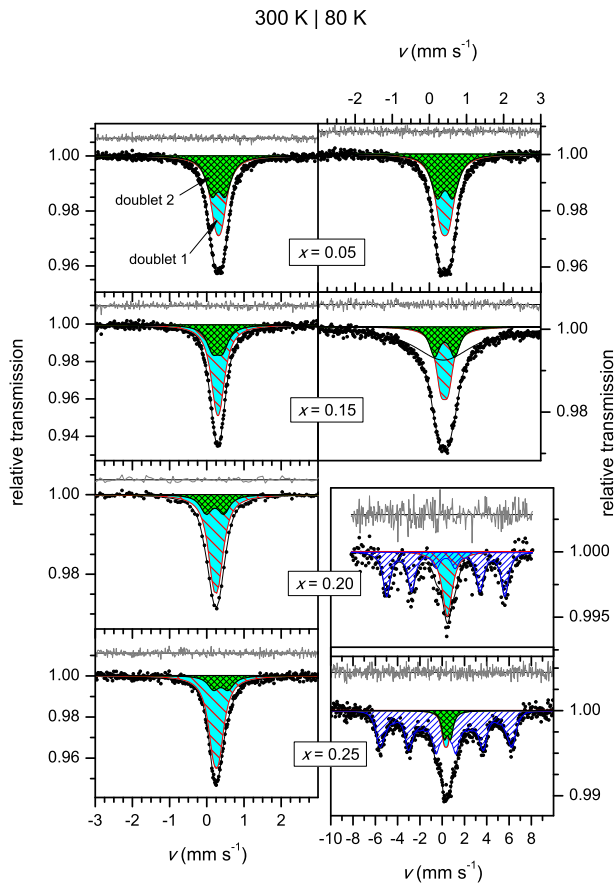}
\caption{\label{fig4} (Color online). $^{57}$Co M\"ossbauer spectra of La$_{1-x}$Sr$_x$CoO$_3$, recorded at room temperature and 77~K (left and right, respectively). Four different Sr$^{2+}$ doping concentrations $x$ are presented (x=0.05, 0.15, 0.2, 0.25). Data for $x$ = 0.2 are taken from Ref. \onlinecite{nemeth2007}. Note the different abscissa scale for the last two panels ($x$ = 0.2 and 0.25 for $T$ = 77~K).}
\end{figure}

In the present room-temperature $^{57}$Co emission M\"ossbauer spectra only one broad resonance line appeared (\textit{cf.} Fig. \ref{fig4}) even in the case of low Sr dopings, which is in contradiction to the observations of Bhide \textit{et al.}\cite{bhide1975prb} Their measurements indicate the existence of another iron species with low isomer shift (around 0~mm/s), which was attributed to low-spin Co$^{3+}$. However, extensive studies in the 1980's revealed that the spin state of the mother cobalt ion cannot be deduced from the nucleogenic iron ion's M\"ossbauer parameters (see the competing acceptors model in Ref. \onlinecite{spiering1990}). Moreover, this value of isomer shift could only be explained assuming low 3d electron density on the transition metals, which is typical for low spin trivalent iron complexes like hexacyanides with high level of orbital hybridization, this is not expected in the case of cobaltate perovskites. So, the nature of that low isomer shift species is rather ambiguous, but large difference in its chemical composition is highly probable.

Fitting the observed M\"ossbauer resonance peaks of the strontium doped compounds with only one doublet component results in FWHM line widths ($\Gamma$) between 0.5 and 0.7~mm/s, which values are unreasonably high compared to the natural lifetime broadening or to the line width of the alpha-$^{57}$Fe:$^{57}$Co calibration spectrum measured with the same PFC absorber ($\Gamma_{cal}$ = 0.32~mm/s). This is a clear indication of overlapping spectral features; therefore, inclusion of a second doublet in the fitting procedure is necessary. The resulting parameters are tabulated in Table \ref{table2}. The two subspectra correspond to the presence of two distinct local microenvironments, \textit{e.g.} two different sites or two different phases. The line widths obtained with the two-component model (0.36-0.39~mm/s) satisfy the expectations, the additional broadening of 0.04-0.07~mm/s being probably due to the random distribution of the environments in these partly disordered samples. The relative area of the fitted doublets shows clear correlation with the Sr$^{2+}$ doping concentration $x$. The isomer shift and quadrupole splitting values primarily reflect the density and spatial symmetry of the valence 3d electrons. Based on the observed values, both doublets for each sample can be identified as those belonging to nucleogenic high-spin trivalent irons. Fitting of the room-temperature M\"ossbauer spectra shows that the two doublets have very similar isomer shift values ($\delta$) for each Sr doping level, however, their quadrupole splittings ($\Delta$) are different.

With lowering the temperature, the shape of the M\"ossbauer spectra depends heavily on the Sr doping level. While the $x$ = 0.05 sample shows no temperature dependence above 77~K, the M\"ossbauer spectrum of La$_{0.85}$Sr$_{0.15}$CoO$_3$ at 77~K has a strong broadening, and the samples with $x$ = 0.2 and 0.25 Sr content present a magnetic phase transition at their corresponding Curie-temperatures (c.f. Fig.\ref{fig1}). However, the presence of the two separated phases persists in the M\"ossbauer spectra for all compositions and temperatures investigated.

\begin{table*}
\caption{\label{table2} Room temperature M\"ossbauer parameters of the La$_{1-x}$Sr$_x$CoO$_3$ perovskites. Line widths were fixed to be the same for the two doublets in each spectrum.}
\begin{ruledtabular}
\begin{tabular}{llllllll}
& \multicolumn{3}{c}{doublet 1}& \multicolumn{3}{c}{doublet 2} & \\ \cline{2-4}	\cline{5-7}
$x$ & area ($\%$) & $\delta$ (mm/s) & $\Delta$ (mm/s) & area ($\%$) & $\delta$ (mm/s) & $\Delta$ (mm/s) & $\Gamma$ (mm/s) \\ \hline
0.05 & 57.8 & 0.316(2) & 0.15(2) & 42.2 & 0.311(3) & 0.41(4) & 0.40(2) \\
0.15 & 68.8 & 0.295(1) & 0.11(2) & 31.2 & 0.292(5) & 0.36(3) & 0.38(1) \\
0.2  & 76.3 & 0.242(2) & 0.15(3) & 23.7 & 0.22(1) & 0.47(7) & 0.37(2) \\
0.25 & 78.2 & 0.255(2)  & 0.13(1) & 21.8 & 0.36(1) & 0.39(1) & 0.38(1) \\
\end{tabular}
\end{ruledtabular}
\end{table*}

\section{Discussion}

\subsection{Phase separation above the magnetic transition temperature}

The theoretical model of cluster formation in manganite perovskites suggested the presence of nanoscale phase separation even at temperatures well above the magnetic phase transition temperature.\cite{kumar2004} This involves that the two phases should be distinguishable at room temperature with local probes. Moreover, small-angle neutron scattering and dc susceptibility measurements suggested that the segregated nanoscale clusters in doped lanthanum cobaltate perovskites exist up to about $T^*\cong$ 350~K (Fig. \ref{fig1}). We have found that this is indeed reflected by the M\"ossbauer spectra, giving a direct spectroscopical evidence for the high temperature MEPS, since two components were needed to fit the room temperature spectra. The two doublet components indicate two different phases, in good accordance with the MEPS model in the corresponding part of the phase diagram. One of the doublets can be related to the Sr$^{2+}$-poor regions, while the other to the Sr$^{2+}$-rich clusters. The doping dependence of the relative area of the doublets helps to identify the two components: the relative area of the one with lower quadrupole splitting (named hereafter as doublet 1, while the one with higher $\Delta$ as doublet 2) increases monotonically with $x$, indicating that it should be related to the Sr$^{2+}$-rich regions (Table \ref{table2}).

The quadrupole splitting ($\Delta$) is a measure of the symmetry of the charge distribution around the probing nucleus; therefore, this parameter can also deliver useful information on the different phases. From the results of the structural characterization (XRD, EXAFS) we saw that the local structure around the cobalt does not change upon the doping, \textit{i.e.} the CoO$_6$ octahedra are not affected (Table \ref{table1}). This holds also for the case of the nucleogenic $^{57}$Fe$^{3+}$, since the electron distribution of the high-spin ion preserves the octahedral symmetry. Therefore, the quadrupole splitting in this case is expected to bring information on the linkage of the CoO$_6$ octahedra. Due to the straightening of Co-O-Co bonds with Sr doping, the asymmetry around the cobalt ions decreases, which agrees well with the observed smaller value of $\Delta$ in the Sr-rich clusters (Table \ref{table2}), providing a further argument for our identification of the phases.

\subsection{Evolution of magnetic phases in the spin glass and ferromagnetic regions}

Concerning the magnetic transitions in LSCO, the M\"ossbauer spectra agrees well with the results of bulk magnetization measurements: both thermal transitions (to spin glass and to unconventional ferromagnet) can be easily identified. In the case of the $x$=0.05 sample the temperature dependence of M\"ossbauer spectra does not show any significant change in the examined 77-300~K temperature interval besides the expected second-order Doppler-shift, which is in accordance with the phase diagram (Fig. \ref{fig1}). In the case of sample $x$=0.15 below the spin glass transition temperature ($T_g$ $\approx$ 100~K) the M\"ossbauer spectra significantly broadens, which indicates increased coupling between the Co magnetic moments; however, the fluctuations are still rapid (as compared to the 142~ns life time of the excited M\"ossbauer probe nucleus), which results in a fast “relaxing” magnetic component (Fig. \ref{fig4}). A rudimentary model can illustrate this case, where the relaxing magnetic component is approximated by a fitted singlet line. The line width of this singlet component is about 1.5~mm/s at 77~K which implies a broadening arising from magnetic relaxation.

\begin{figure}  
\includegraphics[width=8.5cm]{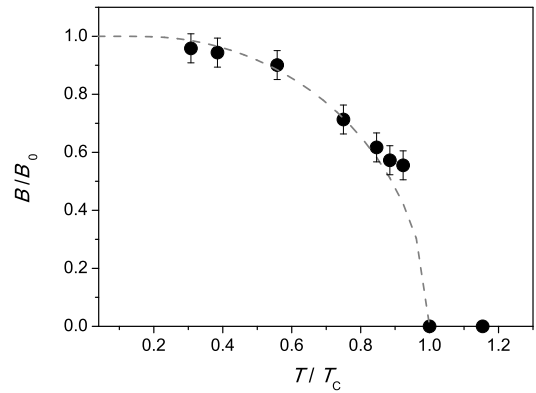}
\caption{\label{fig5} (Color online). Normalized average magnetic hyperfine field deduced from the six-line pattern of the M\"ossbauer spectra of the La$_{0.75}$Sr$_{0.25}$CoO$_3$ sample versus the normalized temperature with $T_C$ = 260~K. The corresponding Brillouin function is illustrated with the dashed line.}
\end{figure}

\begin{figure}  
\includegraphics[width=8.5cm]{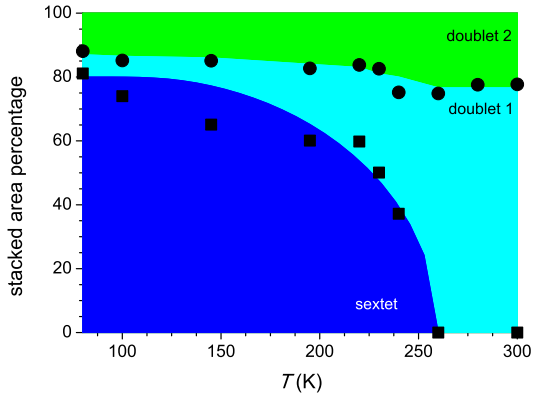}
\caption{\label{fig6} (Color online). Relative area of the three fitted components in the M\"ossbauer spectra of La$_{0.75}$Sr$_{0.25}$CoO$_3$ as a function of temperature. Color fills are guide to the eye. The uncertainty of the values is in the order of the symbol size.}
\end{figure}

In La$_{0.75}$Sr$_{0.25}$CoO$_3$, however, below 260~K the formation of long-range magnetic order is reflected by a well-resolved six-line magnetic pattern that appears beside the two paramagnetic doublets. The temperature-evolution of the ferromagnetic phase can be well reproduced by the Brillouin model, which is fitted the observed magnetic hyperfine field ($B$) deduced from the spectra in Fig. \ref{fig5}, resulting in a $T_C$ of about 260~K (\textit{cf.} Ref. \onlinecite{bhide1975prb}). However, the magnetic transition does not involve all the cobalt sites, since the doublets (although with smaller intensity) are still present even at the lowest measured temperature (77~K). Similar behavior was observed at $x$ = 0.2,\cite{nemeth2007} a composition also above the spin glass to unconventional ferromagnet transition ($x_c\cong$ 0.17)\cite{wu2003}. The coexistence of the doublets and a sextet confirms the separation of ferromagnetic and paramagnetic clusters depicted in the MEPS model. Since the ferromagnetic interaction in these perovskites is originated from the doped electron holes showing a double-exchange-like mechanism, the sextet component can be unambiguously associated with the Sr-rich phase. Moreover, from the evolution of the M\"ossbauer spectra with descending temperature it is clear that the ferromagnetic sextet component develops mainly from doublet 1 (Fig. \ref{fig6}). This again justifies our presumption that doublet 1 can be attributed to the Sr-rich regions.


\subsection{Composition of the doped clusters - microscopic origin of the phase separation}

\begin{figure}  
\includegraphics[width=8cm]{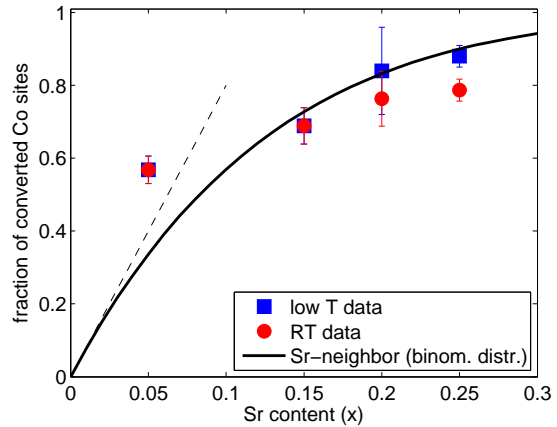}
\caption{\label{fig7} (Color online). Fraction of Co sites converted by doping at low (blue squares) and room temperatures (red circles) as determined from the relative spectral area of the M\"ossbauer component associated with the Sr-rich phase (cf. Table \ref{table2}). Thick solid line: the fraction obtained from binomial distribution for the probability of a cobalt site having at least one neighboring La$^{3+}$ ions replaced by Sr$^{2+}$ out of the 8 possible ones at the given doping level. Dashed line: 8x function representing the scenario where all doping strontium ion converts 8 cobalt ions.}
\end{figure}

Despite of the extensive research in the recent decades, the composition of the separate phases has not been clarified yet, only the terms ``Sr-rich'' and ``Sr-poor'' has been used. However, M\"ossbauer data provide direct quantitative information about the distinct cobalt environments due to the lack of after effects. Irrespective to temperature, the quadrupole splittings of the M\"ossbauer doublets do not change significantly with $x$, only the relative amount of the doublets is affected by the doping. These results are in complete agreement with previous observations on LSCO. The system  is separated into at least two distinguishable phases, and the main variation when increasing the concentration of the dopant is the change in the relative amount of these phases. Moreover, the fraction of Co ions affected by the Sr doping and represented by doublet 1 increases monotonically with \textit{x}, and shows relatively high values (it reaches 90\% even at 25\% Sr concentration). This is in good accordance with neutron diffraction and magnetization studies, which showed that at low doping levels a Sr ion can convert many cobalt ions into higher spin states, leading to a total spin momentum excitation by $\Delta S$=10-16.\cite{yamaguchi1996prbSCO} Moreover, even the first models of DE in manganates assumed that a doping divalent ion introduces a single electron hole that is distributed on all 8 neighboring transition metal ion (or better on the neighboring 8 CoO$_6$ octahedra).\cite{gennes1960pr} These suggest that even one doping Sr$^{2+}$ ion can have a big effect on the valence and spin states of the surrounding cobalt ions. Indeed, assuming a random spatial distribution of the dopants, a simple binominal model, calculating the possibility that a Co has at least 1 Sr neighbor (out of the 8 next La sites), gives a reasonable estimate for the number of converted cobalt ions (Fig. \ref{fig7}). Although the converted cobalt ions cannot be assigned directly to the metallic/ferromagnetic phases due to the complex nature of cluster formation, they should provide the frame of this phase. However, this model insists that the two phases seen by the local M\"ossbauer method can be distinguished as a Sr-free (formally described as ``Sr-poor'') and a Sr-containing (``Sr-rich'') part.

\begin{figure}  
\includegraphics[width=8.5cm]{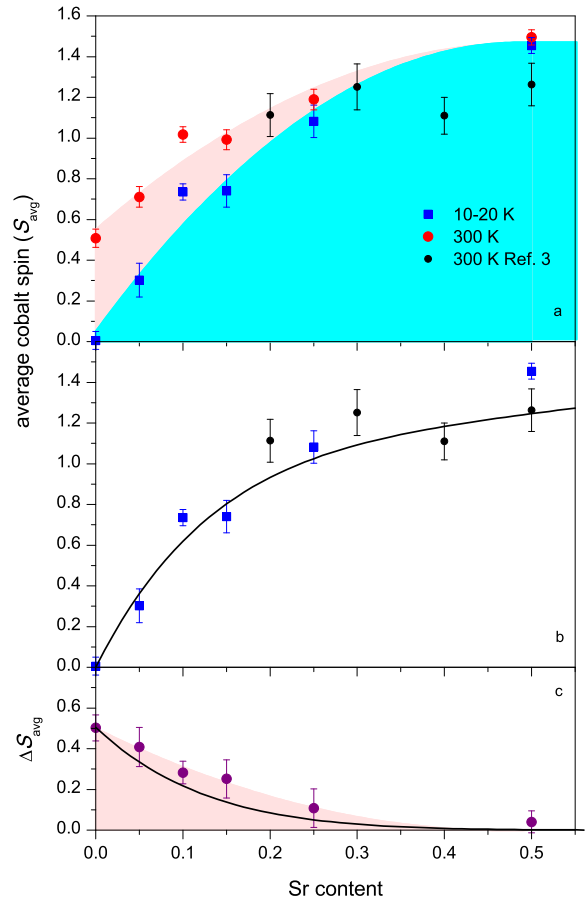}
\caption{\label{fig8} (Color online). (a) Average spin state ($S_{avg}$) of the cobalt ions determined from the XES spectra as a function of temperature and Sr doping. (b) $S_{avg}$ at low temperatures with a binomial distribution function based representation of the spin excitation due to the doping Sr ions (solid line, see text for details). (c) The difference of the 300~K and 10-20~K $S_{avg}$ as a function of Sr content. The solid line shows the scaled ratio of the cobalt ions with no Sr neighbor calculated from binomial distribution. All color fills are guide to the eye.}
\end{figure}

As the spin state of cobalt ions is supposed to be one of the key factors in magnetic interactions, it is important to have a good description of its temperature and doping dependence, for which the bulk-sensitive and element-selective X-ray emission spectroscopy can be advantageously exploited.\cite{vanko2006jpcb} Fig. \ref{fig8}a shows the average spin state of cobalt ions ($S_{avg}$) deduced from the XES spectra with the IAD method \cite{vanko2006prb} for a series of La$_{1-x}$Sr$_{x}$CoO$_3$ ($x$ = 0, 0.05, 0.1, 0.15, 0.25, 0.5). For comparison, the Co spin moments data by Wu \textit{et al.} obtained with bulk magnetization measurements for high ($x$ $\geq$ 0.2) Sr content,\cite{wu2003} is also shown in the figure. As discussed in Section 4.1., the structure of the CoO$_6$ octahedra does not show any significant doping dependence: the oxygen ions surround the cobalt equidistantly with a bond length of about 1.93~\AA. Thus, structural changes on the first oxygen shell cannot be accounted for any doping or temperature driven spin state change. On the other hand, by straightening the Co-O-Co angle, Sr doping may favor higher spin states in cobalt ions.\cite{baier2005} Indeed, up to the percolation threshold ($x_c$ = 0.17), $S_{avg}$ rapidly grows with the doping from the undoped low-temperature value of 0 (all Co$^{3+}$ ions in low-spin state) to about 1. However, the $S_{avg}$ values show a definite change in the slope at $x_c$, growing further to only $S_{avg}$=1.4 at $x$ = 0.5, without any temperature effect. The sudden rise of the average spin state of Co below $x_c$ agrees with the increasing amount of magnetic regions rich in high- or intermediate-spin Co$^{3+}$ ions. The doping Sr ions can convert the neighboring low-spin Co ions to higher spin states. Above the percolation limit the number of the low-spin cobalt neighbors of the doping strontium ions (candidates for spin-state change) decreases suddenly, as many of these Co neighbors are already converted. Thus, the increase rate of the average Co spin state declines. This can be clearly connected to the drastic change of the length scales of the magnetic correlations ($\xi$). Below the critical doping content $\xi$ is smaller than 2-3~nm,\cite{phelan2006} as the converted cobalt ions are rare and far to cooperate. Above $x_c$ the case reverses: the paramagnetic Co$^{3+}$ ions become dense, which is reflected in the moderate $S_{avg}$ grow as well as in the sudden increase of ($\xi$).

From the XES we can separate the thermal and the doping-driven spin transitions. As the applied lowest temperatures ($<$ 20~K) are insufficient to populate higher spin states of cobalt in these perovskites, the low-temperature data should reflect only the effect of strontium doping, while the difference of the low-T and room-temperature values ($\Delta S_{avg}$) corresponds to the thermal spin change. According to our model, the latter part should be proportional to the number of cobalt ions with no Sr$^{2+}$ neighbors. Indeed, the corresponding binomial distribution function, scaled to the thermally excited spin value of LaCoO$_3$ at 300~K, agrees well with the experimental $\Delta S_{avg}$ values (Fig. \ref{fig8}c). Based on this model, the expected low-T spin values of the cobalt ions induced by the Sr doping can be also estimated as follows. The Sr$^{2+}$ doping is expected to create Co$^{3+}$ or Co$^{4+}$ ions with average spin states of $S$=1 and $S$=3/2, respectively.\cite{wu2003} If every doping strontium ion increases the spin value of all eight neighboring cobalt ions with one unit, and also converts one Co$^{3+}$ into Co$^{4+}$, the average spin can be written:

\begin{center}
$S_{avg} = \sum\limits^8_{N=1}{B(N,8,x)\cdot\left[(8-N)\cdot1+N\cdot3/2\right]}$,
\end{center}

where $B$($N$,8,$x$) is the binomial probability density function of finding $N$ out of 8 with a probability of $x$ (here associated with the number of Sr-doped neighbors, number of neighboring La/Sr sites and doping concentration, respectively), while the second term represents the spin-momentum contribution of the excited Co$^{3+}$ and Co$^{4+}$ ions, respectively. Despite being simplistic, this model describes well the observed doping-driven spin state changes (Fig. \ref{fig8}b).
 
The good agreement between the random distribution model of the Sr$^{2+}$ dopants and both M\"ossbauer- and X-ray emission spectra suggests that separated Sr-free and Sr-rich phases can be distinguished. This phase separation is well pronounced even at low doping levels such as $x$ = 0.05, while the spin conversion effect of the Sr doping seems to saturate between 0.15 $<$ $x$ $<$ 0.25. In parallel, the Sr-free phase, decreasing as $B$(0,8,$x$), leads to a vanishing temperature-driven spin excitation. Based on combined neutron diffraction as well as magnetic field dependent heat capacity measurements, it has been proposed that MEPS occurs only in a specific doping rage of 0.04 $<$ $x$ $<$ 0.22.\cite{he2009epl} Moreover, Smith \textit{et al.} have recently reported that, in contrast to earlier phase diagrams,\cite{wu2003} in the inhomogeneous doped material nanoscale hole-poor regions showing spin-state transitions should persist up to the critical doping level $x_c$ = 0.17 in the LSCO perovskites.\cite{smith2012} These converge well with our findings, where we can - with the help of local probe spectroscopic methods - classify the separated phases as Sr-free and Sr-rich, thus refining earlier assumptions, as well as describe the temperature and doping dependency of the evolution of these phases.

\section{Conclusions}

Local X-ray and M\"ossbauer spectroscopies reveal two different atomic environments in the chemically homogenous La$_{1-x}$Sr$_{x}$CoO$_3$ in the doping range of 0.05 $\leq$ $x$ $\leq$ 0.25, as expected for the nanoscale magneto-electronic phase separation, a typical phenomenon in correlated perovskite oxides. These two phases are present even at room temperature, \textit{i.e.}, well above the magnetic transition temperature. One of them rapidly gains weight as the doping level increases, and its local structure is more symmetrical, this phase being identified as the Sr-rich phase. The fraction of this phase agrees with the fraction of the Co ions that have at least one Sr$^{2+}$ neighbors. The spin momentum of Co in this phase increases also rapidly till the percolation threshold, its quantity are well reproduced by the assumption that a Sr$^{2+}$ ion converts all the 8 surrounding Co ions into an average intermediate-spin, which is in perfect agreement with the reports of large spin polarons induced by the Sr$^{2+}$ ions. Accordingly, both the fraction and the spin is described by the same model that invokes random spatial distribution of the Sr$^{2+}$ dopant in the La$^{3+}$ sites, and conversion of the immediate neighbors by the Sr$^{2+}$ ion. Below the Curie temperature the Sr doped phase becomes ferromagnetic. The other phase turns out to be Sr-free, this diminishes rapidly with the increase in the doping level: its fraction is described by the fraction of cobalt atoms that have no Sr$^{2+}$ neighbors. Upon elevating the temperature, this fraction undergoes a spin-state transition, similar to the bulk LaCoO$_3$. Our observations are fully compatible with the nanoscale magneto-electric phase separation model, and it also provides insights into the microscopic origin of this phenomenon in the LSCO system.

\begin{acknowledgments}
This project was supported by the Hungarian Scientific Research Fund (OTKA) under contract No.~K 72597 and the European Research Council via contract ERC-StG-259709. G. Vank\'o acknowledges support from the Bolyai Fellowship of the Hungarian Academy of Sciences.~K. Kn\'{\i}\v{z}ek acknowledges support from the Project No. P204/11/0713 of the Grant Agency of the Czech Republic.
\end{acknowledgments}

\bibliography{c:/munka/cikkek}

\end{document}